# Potential of Traditional Medicinal Plants for Treating Obesity: A Review

Mahnaz Kazemipoor[1], Che Wan Jasimah Wan Mohamed Radzi[2], Geoffrey A. Cordell[3], Iman Yaze[4]

[1,2]Department of Science & Technology Studies, Faculty of Science, University of Malaya, 50603 Kuala Lumpur, Malaysia

[3] Natural Products Inc., Evanston, IL 60203, USA
[4] Faculty of medicine

**Abstract.** Obesity is a global health concern associated with high morbidity and mortality. Therapeutic strategies include synthetic drugs and surgery, which may entail high costs and serious complications. Plant-based medicinal agents offer an alternative approach. A review of the studies on accessible botanical sources for the treatment of obesity is provided, which attempts to explain how these medicinal plants act to cause weight loss, and which approach is safer and more efficient. Information was gathered for the period of 1991 to 2012. Five basic mechanisms, including stimulating thermogenesis, lowering lipogenesis, enhancing lipolysis, suppressing appetite, and decreasing the absorption of lipids may be operating. Consumption of standardized medicinal plant extracts may be a safe treatment for obesity. However, some combinations of medicinal plants may result in either lower efficacy or cause unexpected side-effects.

**Keywords:** Medicinal plants, adipose-tissue differentiation, fat absorption, slimming aids, dietary supplements

## 1. Introduction

By 2005, obesity had affected 400 million adults [66], and since 1997, WHO has cited obesity as a global epidemic [3, 8]. More than half of the adult population in OECD countries is overweight (body mass index [BMI]≥25 $Kg/m^2$) [10]. According to WHO, obesity is related to cardiovascular diseases, hypertension, diabetes mellitus, gallbladder disease, cancer, endocrine and metabolic disturbances, osteoarthritis, gout, pulmonary diseases, as well as psychological issues, including social bias, prejudice, discrimination, and overeating [65]. Economically, obesity and its health consequences place enormous costs now and for future health care [13, 49, 67]. Being overweight is a cosmetic issue, a major health risk factor [33], and may decrease life expectancy [46]. A proliferation of high-cost, anti-obesity products is in the market [25]. However, they exhibit side effects, such as gastrointestinal and kidney problems [25, 55], and only *Orlistat* and *Sibutramine* can be used long-term, in spite of issues regarding weight loss and tolerance [50, 51]. The use of natural remedies for weight loss has increased, based on reliability, safety, and cost compared with synthetic drugs [11] or surgical procedures [12], which may have limitations [40].

### 1.1. Natural medications

Natural ingredients and medicinal plant preparations may enhance satiety, boost metabolism, and speed up weight loss [34, 43]. Including these foods in the diet may therefore assist slow, individual weight loss. However, doubts about human application remain [58]. Despite the global market for satiety, fat burning, dietary supplements and other weight management remedies, patient awareness of these products is insufficient [15]. Here, a brief review of natural medicinal agents and their anti-obesity potential is presented which could aid patients in selecting a botanical product to develop a healthy body.

[+] Corresponding author. Tel.: + (0379675182); fax: + (0379674396).
*E-mail address*: (jasimah@um.edu.my).

## 2. Methods of Data Collection

Data were acquired from various databases, including Science Direct, Pub-Med, Scopus, Web of Science, and from books and theses for the period from 1991 until January 2012. Key search words included: *traditional medicine, medicinal herbs, plant extracts, anti-obesity, weight loss, overweight, botanical remedy, complementary therapy, natural, alternative, phytonutrients, phytochemicals, efficacy, safety, bioactive compounds, appetite, satiety, metabolism, thermogenesis, lipolysis, lipogenesis, adipocytes* and *anthropometric indices*. Papers on human and animal studies, clinical trials, and related to plant-based obesity medication are discussed.

## 3. Results and Discussion

### 3.1 Mechanism of Action

Natural anti-obesity preparations can induce weight loss through several mechanisms. Their functions can be classified into five major categories, as shown in Table 1.

Table1. Different functions of anti-obesity medicinal plants in humans

| No. | Anti-obesity function | Anti-obesity preparations |
|---|---|---|
| 1 | Inhibiting pancreatic lipase activity | chitosan [7, 27], levan [28], mate tea [41], oolong tea [23] jasmine tea [45], green tea [32] |
| 2 | Enhancing thermogenesis | sea weed [36, 38, 39], bitter orange [14, 19, 52], soybean [24] |
| 3 | Preventing adipocyte differentiation | turmeric [1], capsicum [22], palm oil [62], banana leaf [4, 31], brown algae [37], garlic [2], flaxseed [60], black soybean [29] |
| 4 | Enhancing lipid metabolism | herb teas [45], cinnamon [57] |
| 5 | Decreasing appetite | pine nut [48], pomegranate leaf [35], ginseng [30], *Hoodia gordonii* [63] |

Based on pancreatic lipase activity inhibition [5], some medicinal plants will prevent the intestinal lipid absorption, and non-absorbed fat will be excreted through oily faeces. Certain bioactive components can increase the metabolic rate [20] which enhances thermogenesis and helps burning calories and excess body fat. Prevention of adipocyte differentiation [62] through medicinal plant may inhibit adipogenesis and fat cell formation, and based on enhanced lipolysis some medicinal plants can increase lipolysis through inducing β-oxidation or noradrenaline secretion in fat cells [45]. Other anti-obesity ingredients may suppress appetite and induce satiety [16], allowing for appetite control. These differing functions of antiobesity medicinal plants will result in a reduction of food and energy intake [19].

### 3.2 Approaches in medicinal plant preparation with maximum efficacy and safety

Medicinal plant samples may be the whole plant, or plant parts (stem, bark, leaf, flowers, and roots). These materials may be presented as powder or capsules, although the medicinal plants showing antiobesity properties were aqueous or alcoholic extracts. In this way, components inhibiting the anti-obesity compounds may be removed. Extraction, partial purification, and/or the isolation of active principle(s) could increase bioactive constituent bioavailability in the extracts, and enhance medicinal agent efficacy in weight loss [9, 56]. However (Table 2), different antiobesity preparations in combination with other medicinal plants may produce unexpected side-effects.

Table 2 - Comparison of safety and efficacy of single and mixed medicinal plant anti-obesity preparations

| Medicinal Plant ingredient | *Result | Adverse effects | Combination formula | Result | Adverse effects |
|---|---|---|---|---|---|
| Rhubarb (rheum) [26] | + | Not reported | In combination with ginger, astragulus, red sage, turmeric, and gallic acid [18, 53] | greater weight gain in intervention group/ (-) | Musculoskeletal, gastrointestinal, oral, dermatologic, vaginal irritation, headache, etc. |
| Green tea (*Camellia sinensis*) [44] | + ($p<0.05$) | Not reported | In combination with asparagus, black tea, guarana, kidney bean, *Garcinia cambogia* and chromium yeast [47] | no inter group difference in weight (-) | Gastrointestinal complaints |
| Bitter orange *Citrus aurantium*) [59] | + | Not reported | In combination with pantothenic acid, green tea leaf extract, guarana, white willow bark and ginger root [17] | greater weight gain in intervention group ($p<0.04$) | Hypertension, musculoskeletal, neurological, migraine, anxiety |
| Kidney bean (*Phaseolus vulgaris*) [61] | + ($p=0.07$) | Not reported | In combination with green tea extract [6] | inter group diff. (+) | Flatulence, soft stool, constipation |
| *Garcinia cambogia* [42] | + ($p=0.03$) | Not reported | in combination with natural caffeine [54] | No inter group diff. ($p=0.3$) | Not reported |
| Glucomannan fiber [64] | + ($p<0.005$) | Not reported | in combination with chitosan, fenugreek, *Gymemna sylvestre*, vitamin C [68] | inter group differences (+) ($p<0.01$) | Constipation, headache, indigestion |

* Results indicate the efficacy and intergroup differences

The application of single medicinal plants has not caused any adverse events. The undesired human effects may be due to interactions between the different phytochemical constituents present in the different plants [21].

## 4. Summary and Conclusions

Based on a number of *in vivo* studies regarding the efficacy of anti-obesity medicinal plant preparations, they may act by stimulating thermogenesis, lowering lipogenesis, enhancing lipolysis, suppressing appetite, and decreasing lipid absorption. Single and mixed anti-obesity medicinal plant preparations may have different effects. The botanical sources, route of administration, presence of various bioactive components and their respective functions, experimental methods used, treatment dosage, study design, treatment duration, and safety and efficacy of the plant are also factors.

In conclusion, the dietary intake of the single medicinal plants may provide a higher degree of safety and efficacy than mixed medicinal plant preparations. These findings support health organization recommendation regarding the regular consumption of vegetables and selected herbs, such as turmeric, capsaicin, ginger, and green tea. Improving knowledge on the use of anti-obesity medicinal preparations, and encouraging obese patients to consume them along with an enhanced exercise regimen and a healthy diet should be continued. Additional chemical, biological, and clinical studies are needed on the effectiveness of selected medicinal plants, particularly those used as spices and condiments, in ameliorating and treating obesity in humans. Such anti-obesity data would be useful for food and drug manufacturers as new products are developed, and to governments in the regulation of food products as a way to promote and enhance public health.

## 5. Acknowledgments

This review was carried out under Research Grant No. RG108/11SUS, Department of Science & Technology Studies, Faculty of Science, University of Malaya, Kuala Lumpur, Malaysia. The authors are grateful to Ass. Prof. Dr. Dilgit Singh, Faculty of Computer Science and Information Technology for valuable academic guidance.